\documentstyle[epsfig]{qcdparis}
\begin{document}
\makeatletter
\def\stbox#1#2{\hbox{\vtop{\hbox{\strut #1}\hbox{\strut #2}}}}
\def\ps@myheadings{\let\@mkboth\markboth
\def\@oddhead{\stbox{DESY 95-248 \hspace{129mm} ISSN 0418-9833}
                  {December 1995 \hspace{125mm}
hep-ph/9512372
} \hfil}  }
\def\maketitle{\thispagestyle{myheadings}
   \vspace*{2.0cm}
   \begin{center}\@title\end{center}
   \vspace*{1.0cm}
   \normalsize\rm
   \begin{center}\@author\end{center}
   \begin{center}\@address\end{center}
   \vspace*{0.8cm}
   \@collab
   \@abstract
   \vspace*{1.7cm} }
\makeatother
\addtolength{\oddsidemargin}{0.3cm}
\addtolength{\evensidemargin}{0.3cm}
\addtolength{\topmargin}{2.0cm}
\addtolength{\footnotesep}{0.2cm}
\pagestyle{plain}
\thispagestyle{headings}
\title{Direct and Resolved Pomeron in Rapidity Gap Cross Sections$^{\,*\,+}$}
\author{H.-G. Kohrs}
\affil{II. Institut f\"ur Theoretische Physik$^{\ddag}$, Universit\"at
Hamburg\\
and\\
Deutsches Elektronen--Synchrotron DESY, Hamburg, Germany}
\abstract{
  We investigate the effect of a direct pomeron coupling
  to quarks on inclusive jet production
  in DIS and photoproduction. The direct pomeron coupling
  generates a point-like contribution to the diffractive part of
  the structure function $F_2$, which is analysed on the basis of
  the latest H1 and ZEUS data. Our model assumptions for the pomeron
  structure are consistent with the measured data.}
\twocolumn[\maketitle]
\fnm{7}{Talk given in the session of working group I$+$II
at the {\it Workshop on Deep Inelastic Scattering and QCD},
Paris, April 1995, and the Second Meeting of the European Network
{\it Physics at High Energy Colliders}, Como, September 1995}
\fnm{6}{Work done in Collaboration with B.A.~Kniehl and G.~Kramer}
\fnm{2}{Supported by Bundesministerium f\"ur Forschung und Technologie
(BMFT), Bonn, Germany, under Contract 05 6 HH 93P (5) and by EEC
Program {\it Human Capital and Mobility} through Network
{\it Physics at High Energy Colliders} under Contract CHRX-CT93-0357
(DG12 COMA)}
\section{Introduction}
In diffractive production of hadronic final states in $ep$
scattering, the proton stays intact or becomes a low mass state.
Between the direction of the proton remnant, which goes down the
beam pipe, and the produced hadronic system there is no colour
flow, which allows of the possibility to observe large gaps in
rapidity between these directions. The experiments H1~\cite{H1}
and ZEUS~\cite{ZS} at HERA have measured the portion of diffractive
events to be $\approx 10\%$ of all events -- not only in
photoproduction ($Q^2<0.01\,\mbox{GeV}^2$)
but also in deep inelastic scattering (DIS) ($Q^2>10\,\mbox{GeV}^2$).

This paper is organized as follows. First, we describe our ansatz to
analyse diffractive $ep$ scattering. In section 3, we consider the
pomeron structure function and the diffractive part of $F_2$. We find
consistency with the latest HERA data.
Finally, we analyse jet cross sections in diffractive
inclusive photoproduction and in DIS.

\section{Model for diffractive jet production}

There exist various phenomenological models~\cite{IS,La,BH,In}
to describe the above mentioned diffractive
nonperturbative QCD phenomena quantitatively.
We follow a widely spread assumption, where the proton
splits off a colourless object called pomeron (${I\hspace{-1mm}P}$),
which has the quantum numbers of the vacuum. Then, if factorization
holds, the proton vertex can be parametrized by a ${I\hspace{-1mm}P}$--flux
factor that depends on $t=(p-p')^2$, the momentum transfer to the pomeron,
and $x_{I\hspace{-1mm}P}$, the fraction of proton energy,
that it carries away.

In fact, this has been done in the past by various authors, who
fixed their parameters with the help of $p\overline{p}$ scattering data.
Inspired by Regge phenomenology, Berger et al.~\cite{BCSS} found
for the pomeron flux
\begin{eqnarray}
\label{Berger}
      f_{I\hspace{-1mm}P/p}(t,x_{I\hspace{-1mm}P})
      \hspace{-2mm}&=&\hspace{-2mm}
      \frac{\beta_{I\hspace{-1mm}Pp}^2(t)}{16\pi}
      x_{I\hspace{-1mm}P}^{1-2\alpha(t)}
\end{eqnarray}
with the Regge trajectory $\alpha=\alpha_0+\alpha't$ and the
residue function
$\beta_{I\hspace{-1mm}Pp}^2(t)=\beta_{I\hspace{-1mm}Pp}^2(0) e^{b_0t}$,
where $\alpha_0=1+\epsilon$, $\epsilon=0.085$, $\alpha'=0.25\,\mbox{GeV}^{-2}$,
$\beta_{I\hspace{-1mm}Pp}^2(0)=58.74\,\mbox{GeV}^{-2}$
and $b_0=4.0\,\mbox{GeV}^{-2}$.

This definition of the factorization in a pomeron
flux factor differs by a factor of $\frac{\pi}{2}$ from the definition
of Donnachie and Landshoff~\cite{DL}. In addition, they
included the Dirac elastic form factor of the proton, which is
given by
\begin{eqnarray}
F_1(t)=\frac{(4m_p^2-2.8t)}{(4m_p^2-t)}\frac{1}{(1-t/0.7)^2}
\end{eqnarray}
and yielded
\begin{eqnarray}
\label{fDL}
      f_{I\hspace{-1mm}P/p}(t,x_{I\hspace{-1mm}P})
      \hspace{-2mm}&=&\hspace{-2mm}
      \frac{9\delta^2}{4\pi^2} [F_1(t)]^2
      x_{I\hspace{-1mm}P}^{1-2\alpha(t)}
\end{eqnarray}
with $\delta=3.24\,\mbox{GeV}^{-2}$.
Equation \ref{fDL} may be regarded as the most natural way
to define a flux factor \cite{La} and will be used in the following.
The same definition of the flux factor was used by Ingelman and
Schlein~\cite{IS}. They made a different parametrization with
two exponentials of the following form
\begin{eqnarray}
      f_{I\hspace{-1mm}P/p}(t,x_{I\hspace{-1mm}P})
      \hspace{-2mm}&=&\hspace{-2mm}
      \frac{1}{2} \frac{1}{\kappa x_{I\hspace{-1mm}P}}
      (A\exp^{\alpha t} + B\exp^{\beta' t})
\end{eqnarray}
with the parameters $\kappa=2.3\,\mbox{GeV}^2$, $A=6.38$,
$\alpha=8\,\mbox{GeV}^{-2}$, $B=0.424$ and $\beta'=3\,\mbox{GeV}^{-2}$.
Here, the factor of $1/2$ comes in because of
normalization to {\it one} proton vertex.

For our purposes, the momentum transfer $t=(p-p')^2$ to the
pomeron has to be integrated out, since the proton remnant is
(still) not tagged. We get
\begin{eqnarray}
\label{flux}
      G_{I\hspace{-1mm}P/p}(x_{I\hspace{-1mm}P})
      \hspace{-2mm}&=&\hspace{-2mm}
      \int_{-\infty}^{t_2} dt\,
      f_{I\hspace{-1mm}P/p}(t,x_{I\hspace{-1mm}P})
\end{eqnarray}
\begin{eqnarray}
      \mbox{with} \hspace{15mm}
      t_2=-m_p^2 x_{I\hspace{-1mm}P}^2/(1-x_{I\hspace{-1mm}P})
      \quad.
      \nonumber
\end{eqnarray}

We emphasize that recently published H1 data~\cite{H1} and ZEUS
data~\cite{ZS} confirm this $x^{-n}$ factorization.
The actual values for the exponent  $n$ are
\begin{eqnarray}
\label{exponent}
      n &=& 1.19\pm 0.06\pm 0.07 \quad \mbox{H1 \cite{H1}}  \quad, \\
      n &=& 1.30\pm 0.08_{-0.14}^{+0.08} \quad \mbox{ZEUS \cite{ZS}}
      \quad ,
\end{eqnarray}
which are comparable to the exponent $2\alpha(0)-1=1.17$ from Regge
analysis. However, since the pomeron
might not be a real particle, there could be problems with the
interpretation of (\ref{flux})~\cite{La}.

Surely, the pomeron is in some sense only a generic object, that serves
to parametrize a nonperturbative QCD effect. Although it is not
considered as a physical particle that could be produced in the
s--channel, we employ the concept of structure functions for it.

For the hadron--like part of the unknown parton density functions
of the pomeron, $G_{b/I\hspace{-1mm}P}$, we propose the ansatz
\begin{eqnarray}
\label{distri}
      \beta\,G_{u/I\hspace{-1mm}P}(\beta)
      \hspace{-2mm}&=&\hspace{-2mm}
      \beta\,G_{\overline{u}/I\hspace{-1mm}P}(\beta)
      =\beta\,G_{d/I\hspace{-1mm}P}(\beta)
      =\beta\,G_{\overline{d}/I\hspace{-1mm}P}(\beta) \nonumber\\
      \hspace{-2mm}&=&\hspace{-2mm}
      6\beta(1-\beta) \frac{1}{5} \frac{1}{1+r} \quad,\\
      \beta\,G_{s/I\hspace{-1mm}P}(\beta)
      \hspace{-2mm}&=&\hspace{-2mm}
      \beta\,G_{\overline{s}/I\hspace{-1mm}P}(\beta)
      =\frac{1}{2}\beta\,G_{u/I\hspace{-1mm}P}(\beta) \quad,\nonumber\\
      \beta\,G_{g/I\hspace{-1mm}P}(\beta)
      \hspace{-2mm}&=&\hspace{-2mm}
      6\beta(1-\beta) \frac{r}{1+r}
      \nonumber
\end{eqnarray}
and vanishing charm contributions. Here, $\beta=x/x_{I\hspace{-1mm}P}$
with Bjorken $x$. These functions obey the sum
rule
\begin{eqnarray}
\label{sum}
  \sum_b\int_0^1 d\beta\,\beta\,G_{b/I\hspace{-1mm}P}(\beta)
  &=& 1
\end{eqnarray}
and are, in our case, defined for an input scale of
$Q_0^2=2.25\,\mbox{GeV}^2$.
Theoretical motivation on the basis of nonperturbative
methods can be found in \cite{BCSS} and \cite{BS}.
The analysis of the authors suggests a behaviour
of the diffractive parton distributions between $(1-\beta)^0$
and $(1-\beta)^2$ for $\beta \rightarrow 1$.

The parameter $r$ describes the unknown ratio of the
gluon to quark content of the pomeron. To get a first
insight into the structure of our model, we
restrict the number of free parameters and choose a value $r=3$,
which represents simple gluon dominance in the pomeron.
We carry out the usual DGLAP--evolution to get the right $Q^2$
dependence of these functions~\cite{Vo}.

Several groups considered the possibility of a direct pomeron
coupling to quarks \cite{DoLa,CFS,BS2,CHPWW}. A direct pomeron
coupling corresponds to a $\delta$--function term in the
pomeron structure function and produces a leading--twist
behaviour in the $p_T$ spectrum. Our purpose is to find
criteria that allow us to see a direct pomeron coupling in the
data~\cite{KKK}.

As a consequence, similarly to $\gamma\gamma$ scattering,
the $\gamma I\hspace{-1mm}P \rightarrow q\overline{q}$ cross
section also contributes to the pomeron structure function. Here,
we assume a direct vector coupling of the pomeron to the quarks
with coupling strength $c$. This is not really justified with
respect to the $C$ parity. However, the $Q^2$ dependence, that is $\sim
\log Q^2$ at low $x$, is only weakly dependent on the spin structure.
This can be seen, for instance, if one replaces the vector coupling
by a scalar one. To remove the collinear singularity, we introduce
the regulator quark masses $m_q$ and obtain for the point--like (pl)
part
\begin{eqnarray}
\label{point}
    \hspace{-2mm}&&\hspace{-2mm}
    \hspace{-5mm} \beta\,G_{q/I\hspace{-1mm}P}^{pl}(\beta,Q^2)
    \nonumber\\
    \hspace{-2mm}&&\hspace{-8mm}
    =\,\beta\,\frac{N_c}{8\pi^2}\,c^2 \Bigg\{v
    \bigg[-1+8\beta\left(1-\beta\right)
    -\frac{4m_q^2}{Q^2}\beta\left(1-\beta\right)
    \bigg] \nonumber\\
&&
    \hspace{-8mm} +\bigg[\beta^2+\left(1-\beta\right)^2
    +\frac{4m_q^2}{Q^2}\beta\left(1-3\beta\right)
            -\frac{8m_q^4}{Q^4}\beta^2 \bigg]
    \ln\frac{1+v}{1-v}\Bigg\}
    \nonumber
\end{eqnarray}
\begin{eqnarray}
\mbox{with} \hspace{15mm}
    v \hspace{-2mm}&=&\hspace{-2mm}
    \sqrt{1-\frac{4m_q^2\beta}{Q^2(1-\beta)}} \quad.
\end{eqnarray}
\ffig{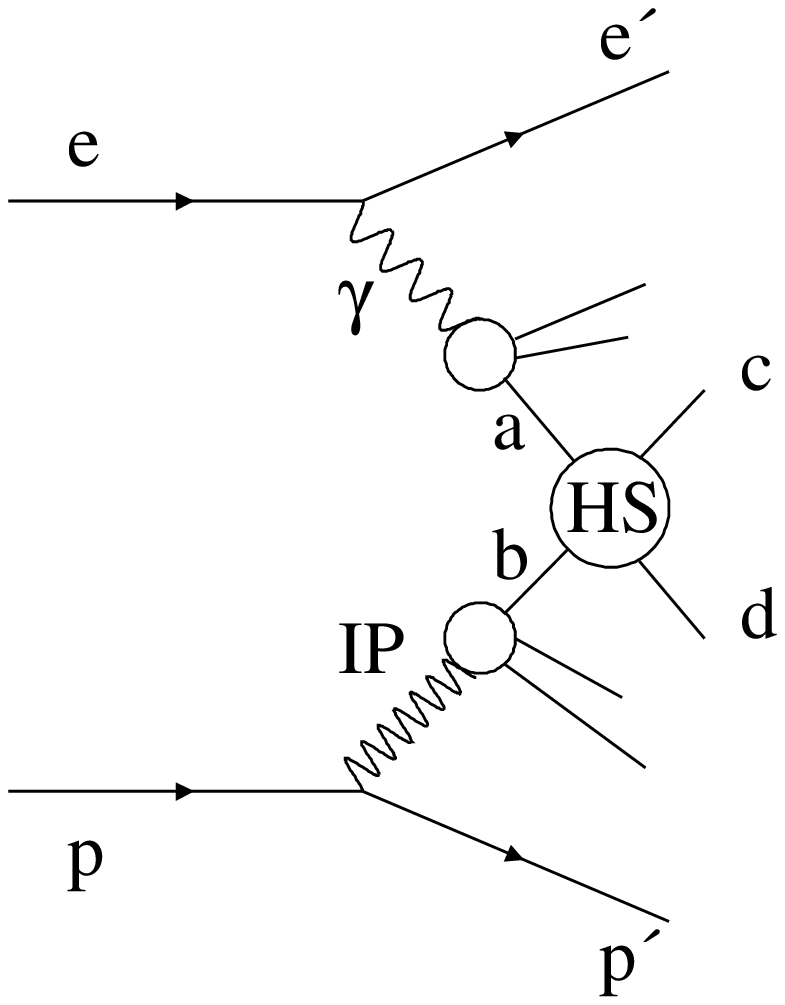}{50mm}{\em Generic diagram for the diffractive $ep$
scattering process with a resolved photon $\gamma$, a resolved
pomeron ${I\hspace{-1mm}P}$ and the hard subprocess $HS$.
}{fig1}
\ffig{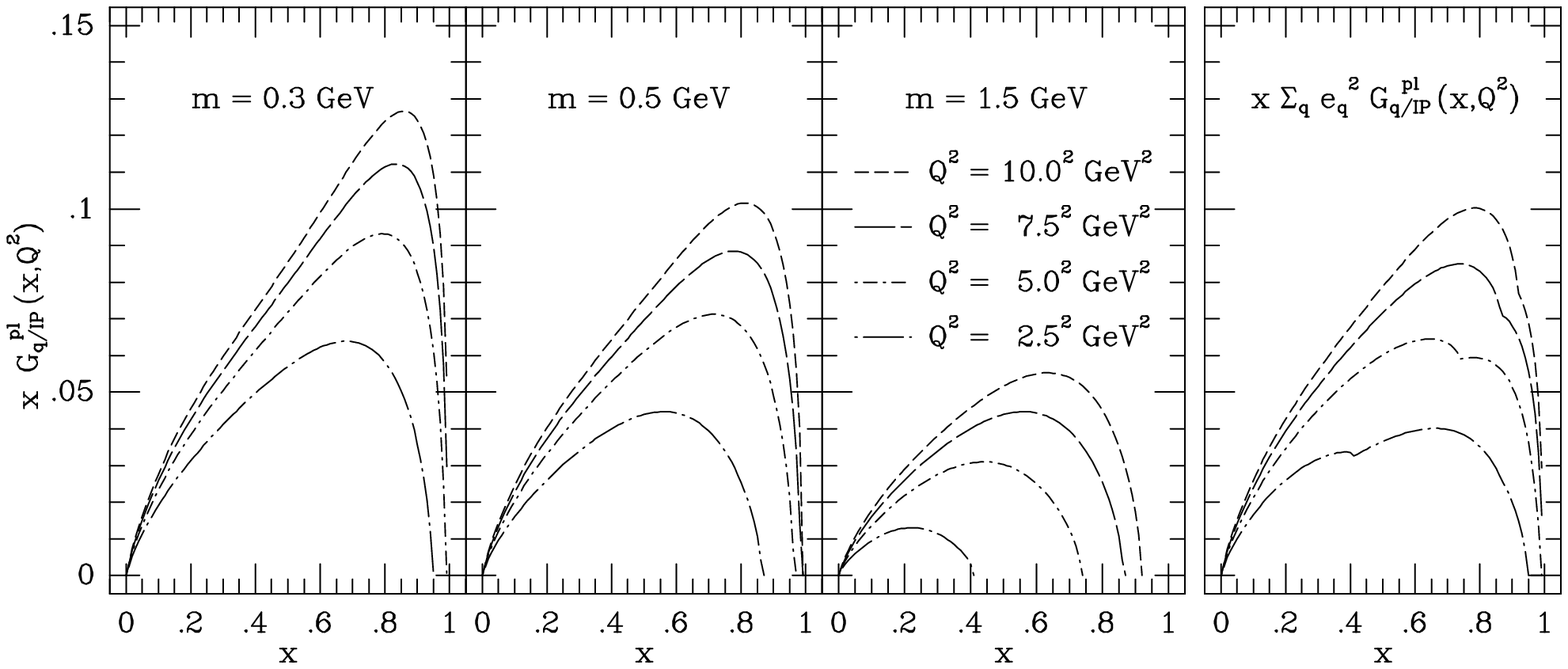}{45mm}{\em The pl distribution function for various
$Q^2$ values and our choice of regulator quark masses. The right
picture shows the $e_q^2$ weighted sum that enters into
$F_2^D$. In this plot $x$ is the variable $\beta$ in
(\protect\ref{point}).
}{fig2}
The point--like contribution for the three quark masses
$m_u=m_d=0.3\,\mbox{GeV}$, $m_s=0.5\,\mbox{GeV}$ and
$m_c=1.5\,\mbox{GeV}$ is plotted in \fref{fig2}. The right
picture shows the $e_q^2$--weighted sum of the pl contributions from
$u$, $d$, $s$ and $c$ quarks to $F_2^D$. The dents in the curves
are caused by the charm threshold. At fixed regulator mass,
the maximum of the point--like contribution is shifted towards
$x=1$ with increasing scale $Q^2$.

To get a constraint on the coupling $c$ under the condition
$r=3$, we analyse the recently published data of $F_2^D$,
the diffractive part of the proton structure function.

\section{The diffractive contribution to $F_2$}

To leading order in $\alpha_s$, only the quark distributions of the
pomeron enter into the deep--inelastic ${I\hspace{-1mm}P}$
structure function $F_2^{I\hspace{-1mm}P}(\beta,Q^2)$, which is
\begin{eqnarray}
\label{F2IP}
    \hspace{-2mm}&&\hspace{-2mm}
    \hspace{-16mm} F_2^{{I\hspace{-1mm}P}}(\beta,Q^2) \\
      \hspace{-2mm}&=&\hspace{-2mm}
    \sum_q
    e_q^2\,\beta\,\left[G_{q/{I\hspace{-1mm}P}}(\beta,Q^2)
    + G_{\overline{q}/{I\hspace{-1mm}P}}(\beta,Q^2) \right.\nonumber\\
      \hspace{-2mm}&&\hspace{-2mm}
    \left.
    + 2\,G_{q/{I\hspace{-1mm}P}}^{pl}(\beta,Q^2)\right]
    \nonumber\quad.
\end{eqnarray}
The comparison with preliminary 1993 H1 data~\cite{H1}
and ZEUS data~\cite{ZS} is shown in  \fref{fig3} and \fref{fig4}.
If factorization holds, which is favoured by
the experiments for a large range of $\beta$ and $Q^2$ values,
the data points are proportional to
$F_2^{{I\hspace{-1mm}P}}(\beta,Q^2)$, i.e.
$\tilde{F}_2(\beta,Q^2)=k\,F_2^{{I\hspace{-1mm}P}}(\beta,Q^2)$
with a constant $k$ that is determined by the
${I\hspace{-1mm}P}$-flux factor:
\begin{eqnarray}
    k &=&
    \int_{x_{I\hspace{-1mm}P}^{min}}^{x_{I\hspace{-1mm}P}^{max}}
    dx_{I\hspace{-1mm}P}\,\int_{t_1}^{t_2} dt\,
    f_{{I\hspace{-1mm}P}/p}(x_{I\hspace{-1mm}P},t)
    \quad.
\end{eqnarray}
The absolute normalization due to $k$ is very sensitive to the
integration bounds, $x_{I\hspace{-1mm}P}^{min}$ and
$x_{I\hspace{-1mm}P}^{max}$, which are taken from the
respective experiments (see \fref{fig3}
and \fref{fig4}). Unfortunately,
no experimental errors on them have been published yet.
A small reduction of the integration interval
would improve our normalization to the data
substantially.

We concentrate therefore on the
discussion of the shapes of the data in comparison to our model.
We find that our model fits the shape of the data curves well.
Especially, the $Q^2$--evolution is fitted better for a combined
ansatz, i.e., quarks in the pomeron {\it and} pl part (solid
curves in \fref{fig3}), than for the DGLAP evolved quark
distributions (short dashed line in \fref{fig3}) or pl part alone.
An alternative possibility to fit the data has been represented in
\cite{ZS}.
\ffig{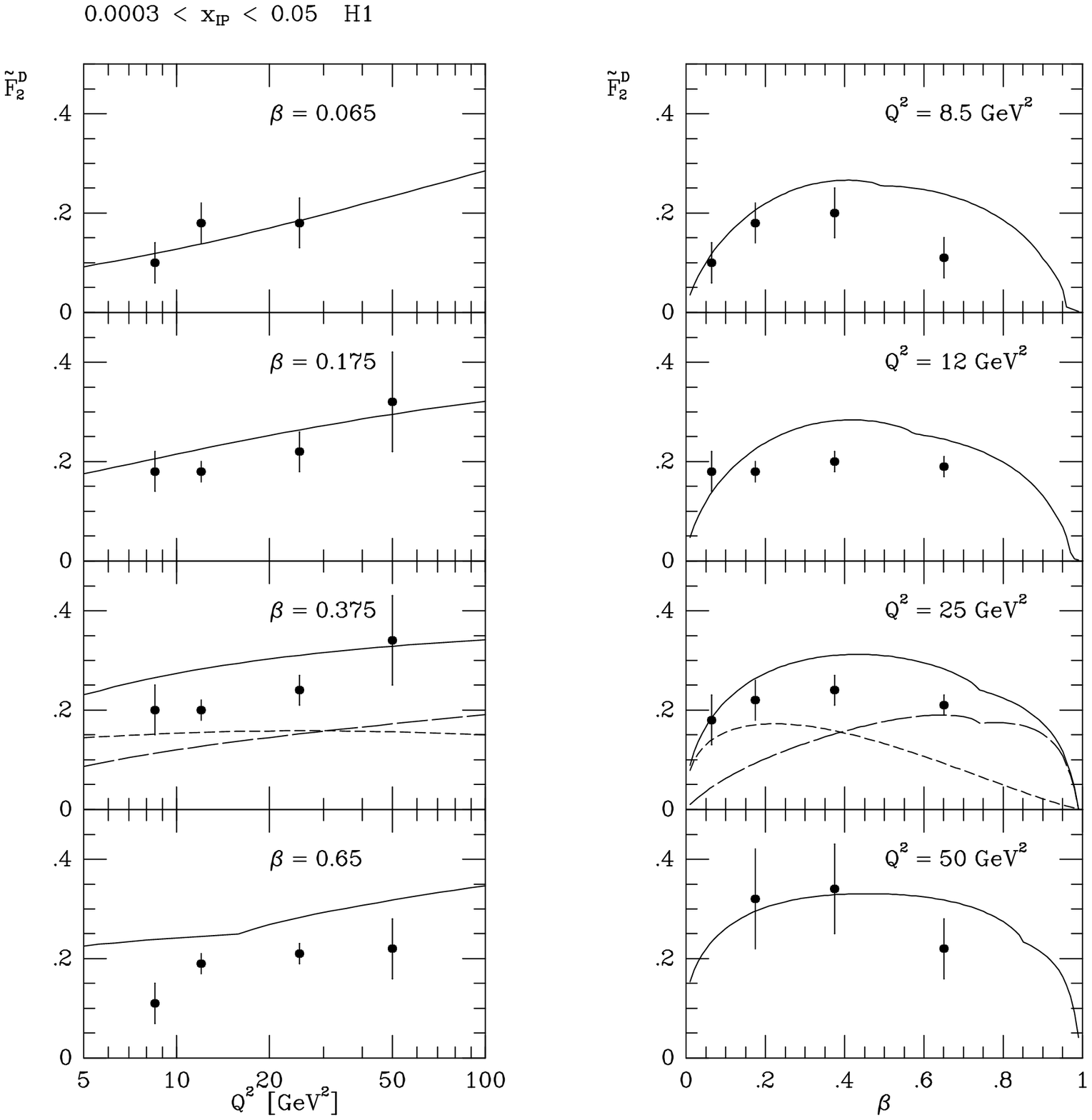}{92mm}{\em Comparison of our model predictions with
preliminary 1993 H1 data \protect\cite{H1}. An important feature
of the point--like
contribution (long--dashed line) is the filling up of
the quark distributions (short--dashed line) at higher $\beta$
where they become less dominant. This results in flatter curves
for the combined distribution (full line) and a better fit to
the shape of the data.
}{fig3}

\ffig{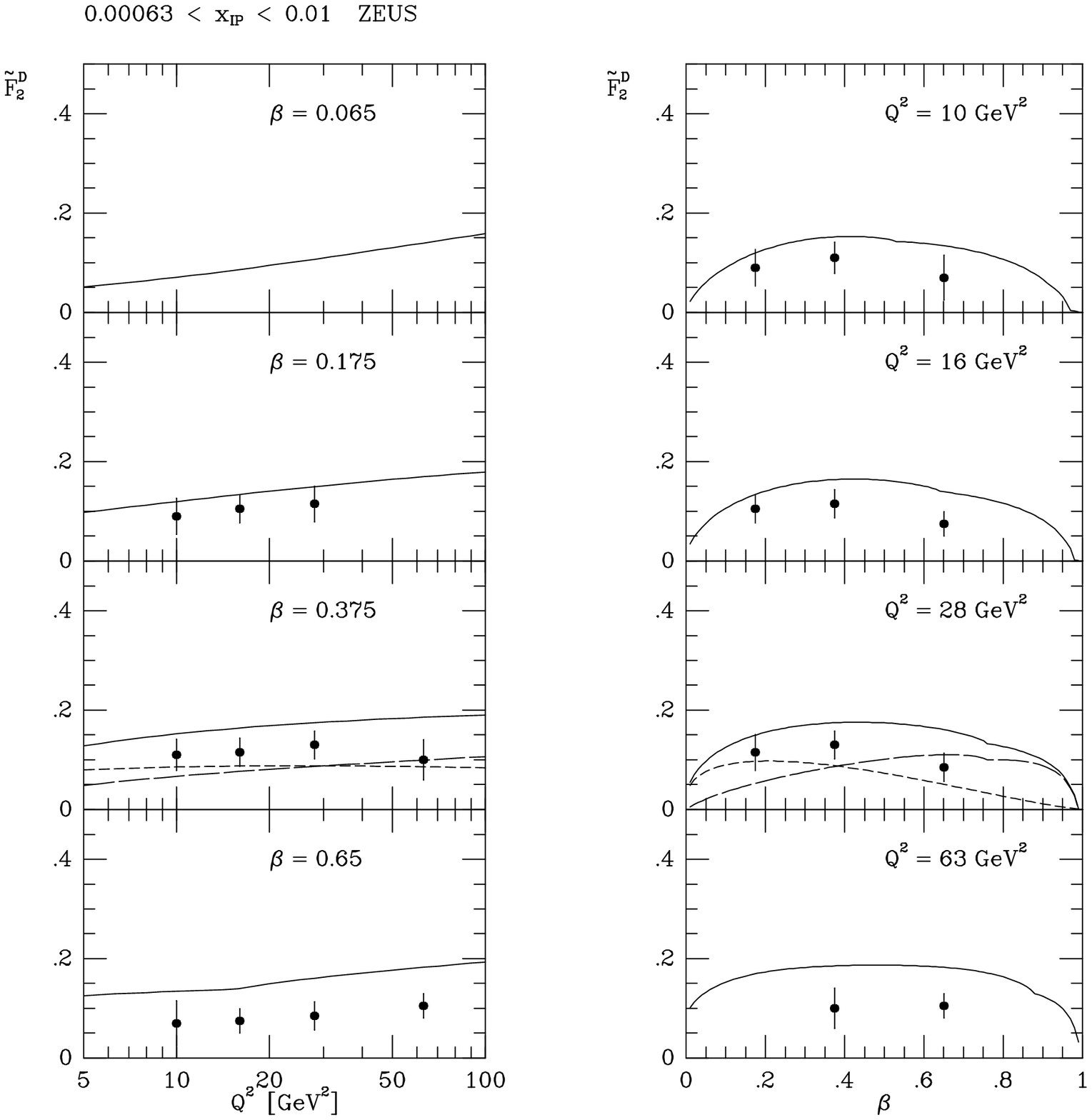}{92mm}{\em Comparison of our model predictions with
preliminary ZEUS data \protect\cite{ZS}, analogous to \protect\fref{fig3}.
}{fig4}

Finally, we fold the pomeron structure function (\ref{F2IP}) with the
pomeron flux factor in eq.~(\ref{flux}) to get the diffractive part
of the proton deep--inelastic structure function $F_2^D(x,Q^2)$.
The relation is
\begin{eqnarray}
\label{F2D}
    \hspace{-2mm}&&\hspace{-2mm}
    \hspace{-10mm} F_2^{D}(x,Q^2) \\
    \hspace{-2mm}&=&\hspace{-2mm}
    \int_x^{x_0}
    dx_{I\hspace{-1mm}P}\,\int_{t_1}^{t_2} dt\,
    f_{{I\hspace{-1mm}P}/p}(x_{I\hspace{-1mm}P},t)
    F_2^{I\hspace{-1mm}P}(x/x_{I\hspace{-1mm}P},Q^2) \quad.
    \nonumber
\end{eqnarray}
The upper bound $x_0=0.01$ is an experimental choice.
In contrast to the analysis of the pomeron structure function, the variable
Bjorken $x$ is now fixed (instead of $\beta$). In \fref{fig5}, we compare
with 1993 H1 data~\cite{H1006}. Again, we emphasize the consistency of the
choice $c=1$ ($r=3$) for the direct pomeron coupling in our model
and the data. We use this value in the calculation of the diffractive
jet cross sections.
\ffig{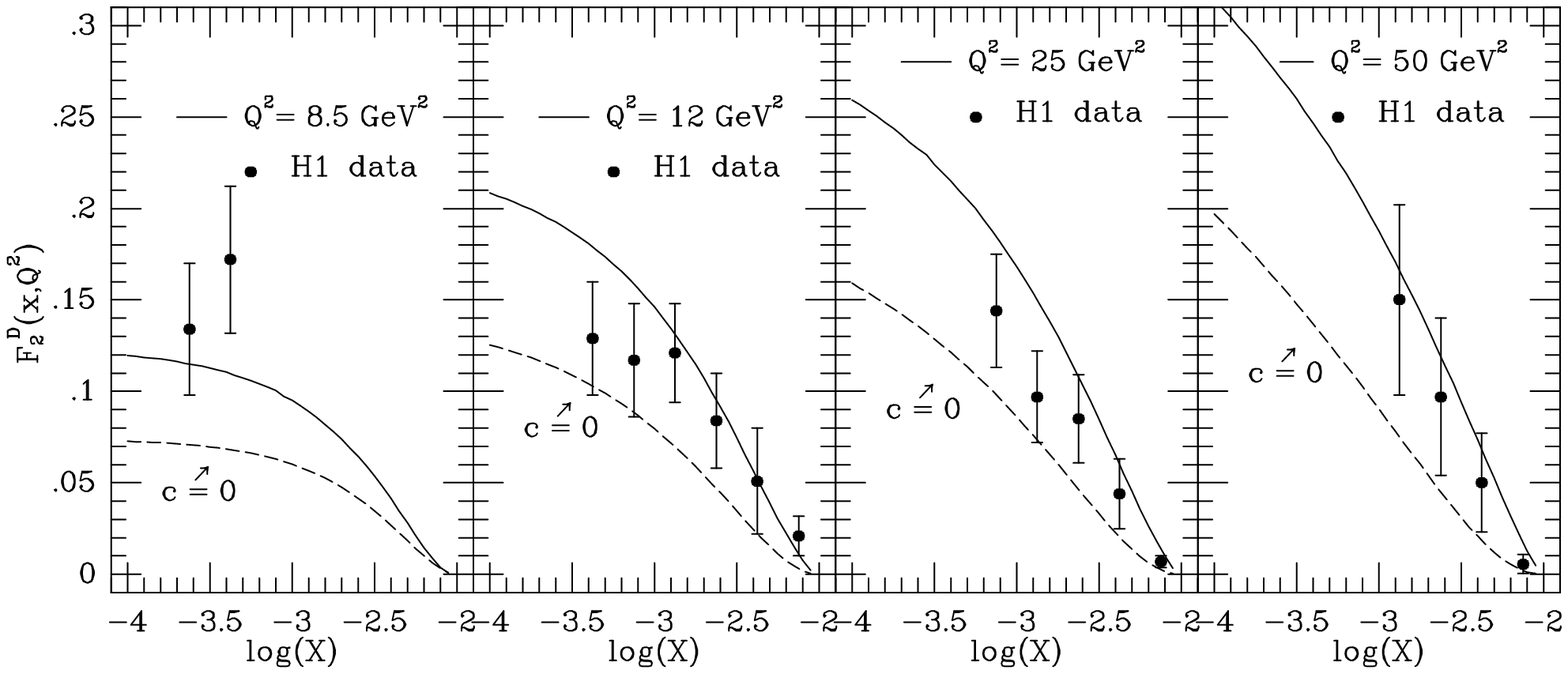}{45mm}{\em $F_2^D(x,Q^2)$ compared to preliminary
1993 H1 data \protect\cite{H1006}. The dashed lines represent
only the contributions from quarks in the resolved pomeron,
while for the solid curves the pl contribution with $c=1$
is included.
}{fig5}
\section{Jets in diffractive inclusive photoproduction and DIS}
The calculation of the differential jet cross section
$\frac{d^2\sigma}{dy dp_T^2}$ of the process depicted in
\fref{fig1} is straightforward. Here, $p_T$ and $y$ are the
transverse momentum and rapidity of one outgoing jet.
In the case of photoproduction, we perform
the evaluation in the $ep$ laboratory system where the rapidity
is positive for jets travelling in the proton direction.
We have then the usual factorization of the photon
flux factor at the electron vertex~\cite{WW}. As is well known
in photoproduction, the photon is resolved or couples directly
to the final--state quarks. For the photon particle density
functions, we take the parametrizations of GRV~\cite{GRV}.

\ffig{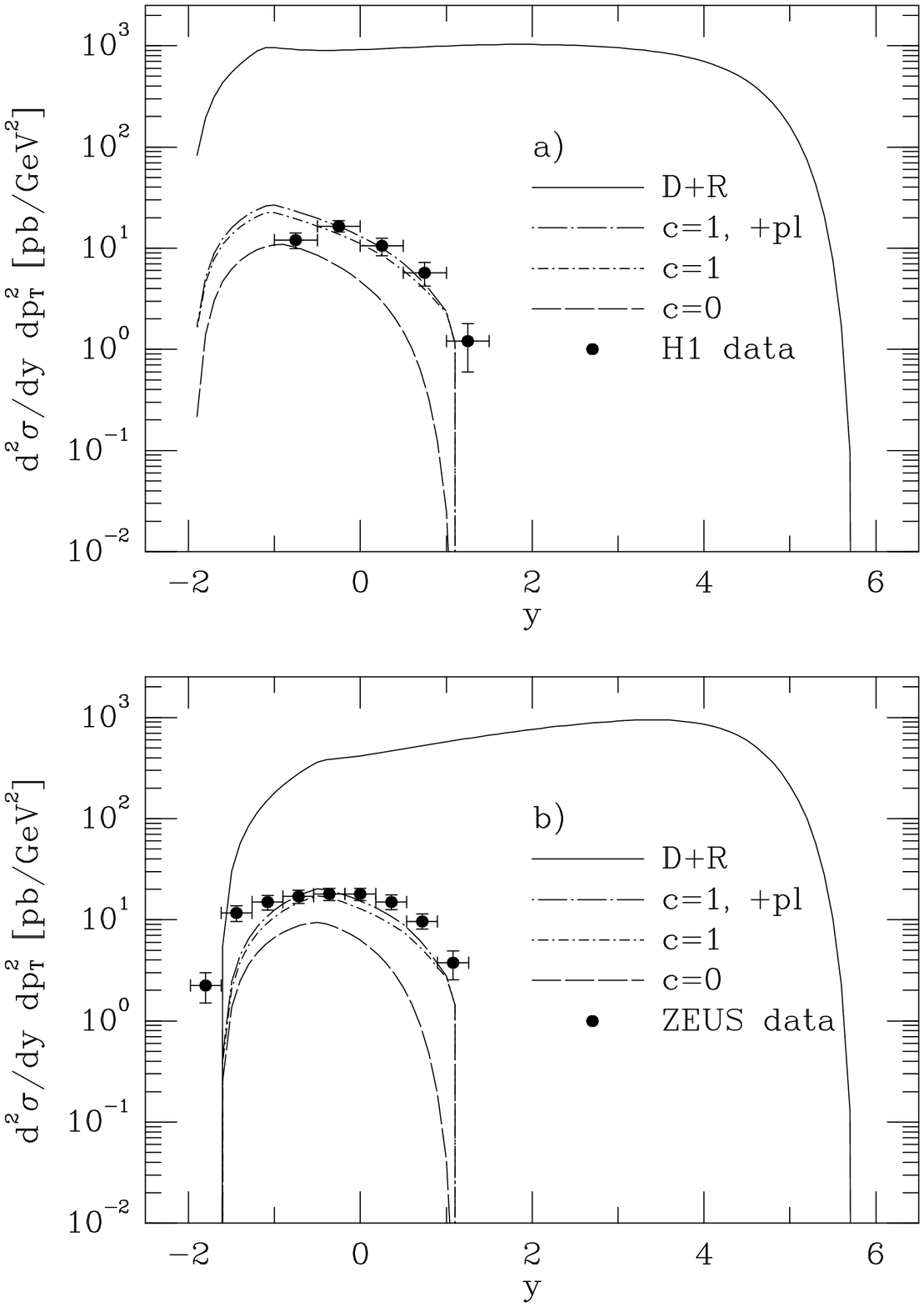}{120mm}{\em
The rapidity distribution of the a) one-- and b) two--jet cross
sections for fixed transverse momentum $p_T=5\,\mbox{GeV}$ in the
$ep$ laboratory system. Here, $y$ is defined to be positive for
jets travelling in the proton direction. For comparison, the
nondiffractive cross section obtained with CTEQ parametrizations
of the proton structure functions is also shown (solid line).
Since the 1993 event rates (data points) of H1 \protect\cite{H1112}
and ZEUS \protect\cite{ZS210} are not normalized
to the luminosity, we can compare only the shapes.
}{fig6}

\ffig{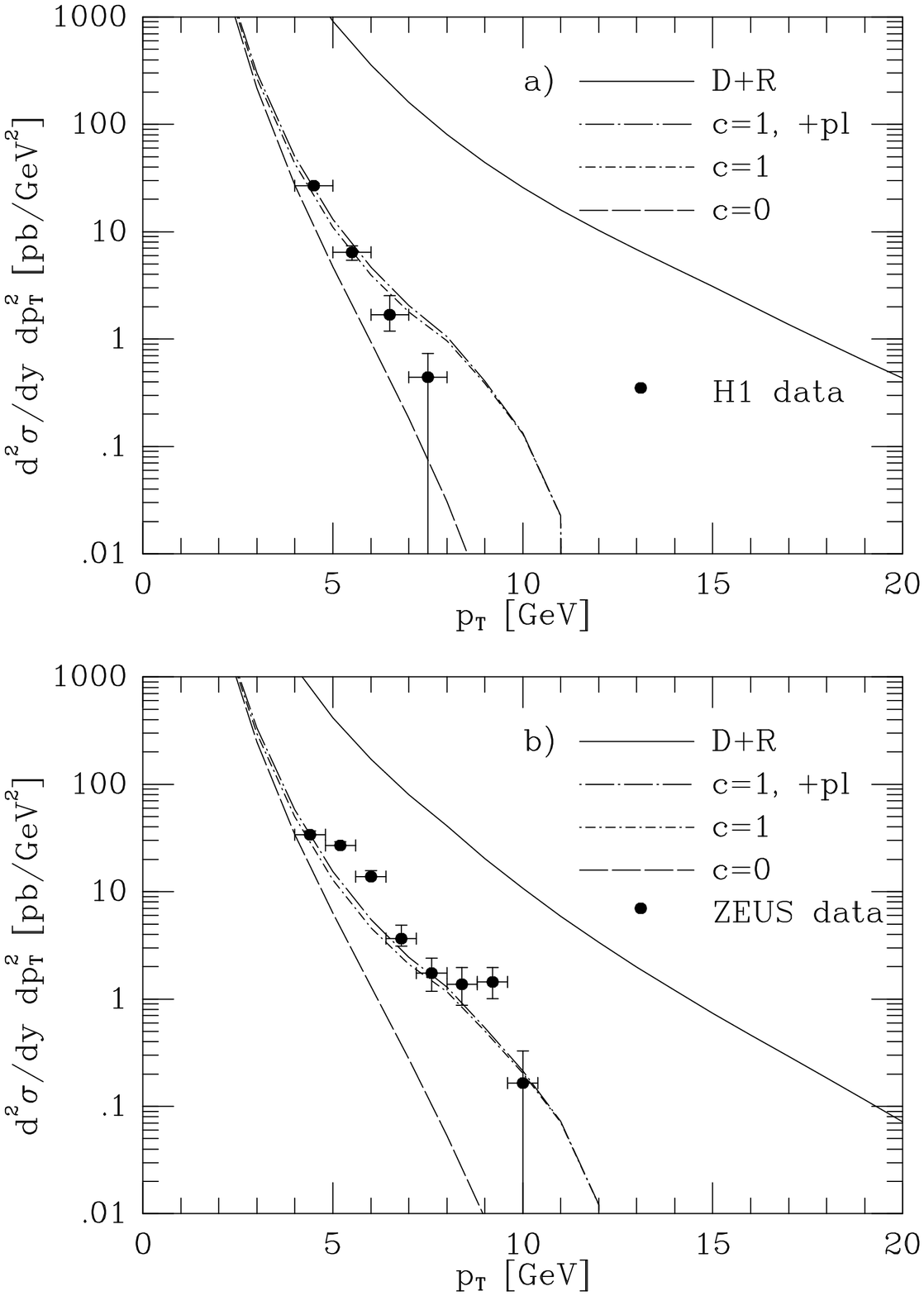}{120mm}{\em
The $p_T$ spectra of the a) one--jet and b) two--jet cross sections
for fixed rapidity $y=0$. Like in \protect\fref{fig6}, the
nondiffractive cross section obtained with CTEQ parametrizations
of the proton structure functions is also shown (solid line).
Since the 1993 event rates (data points) of H1 \protect\cite{H1215}
and ZEUS \protect\cite{ZS210} are not normalized
to the luminosity, we can compare only the shapes. For higher
$p_T$ values the contribution from a direct pomeron quark coupling
dominates the resolved part (c=0) and is favoured by the shape of
the data.
}{fig7}

A more detailed discussion can be found in~\cite{KKK}. However, here
we have included the improved quark distributions of the pomeron due to
the performed $Q^2$--evolution with $Q=p_T$, the transverse momentum of
the considered jet. Further, we use in our analysis more actual data of the
run in 1993.

The differential inclusive one--jet cross section is obtained
by integrating out all kinematic variables over the allowed ranges
without regard to the rapidity of the second jet, while for the
two--jet cross section, we demand explicitly that the second jet
does not enter the cone that is set up by the first jet
around the outgoing proton direction.

The results are shown in \fref{fig6}. Note the large
rapidity gap in the forward direction between the
diffractive and nondiffractive parts. In our analysis
this is controlled by the $x_{I\hspace{-1mm}P}^{max}=0.01$ cut.
This corresponds to a $y^{max}\approx 1.2$ and can
be compared to the experimental value $y^{max}_{exp}\approx 1.5$.
The lower limit in the $y$--distribution is due to the cut
on $x_\gamma$ that enters the improved Weizs\"acker Williams formula
and is numerically determined by the experimental conditions.

The need of a direct pomeron coupling becomes clear, if one
inspects the slope of the $p_T$ spectra in \fref{fig7}.
As expected, with the direct coupling, the $p_T$ spectrum does not fall
off so strongly compared to the resolved pomeron contribution ($c=0$).
The point--like component of the structure function, however,
does not play a significant role in this discussion, which can be
understood by the gluon dominance in our model ($r=3$ in eqs. (\ref{distri})).

A reduction of $r$ would increase the quark content in the
pomeron due to the sum rule, eq. (\ref{sum}). But this would
be accompanied by a reduction of the coupling $c$ to
satisfy the bounds coming from the analysis of the diffractive
part of $F_2$ in section 3.

\ffig{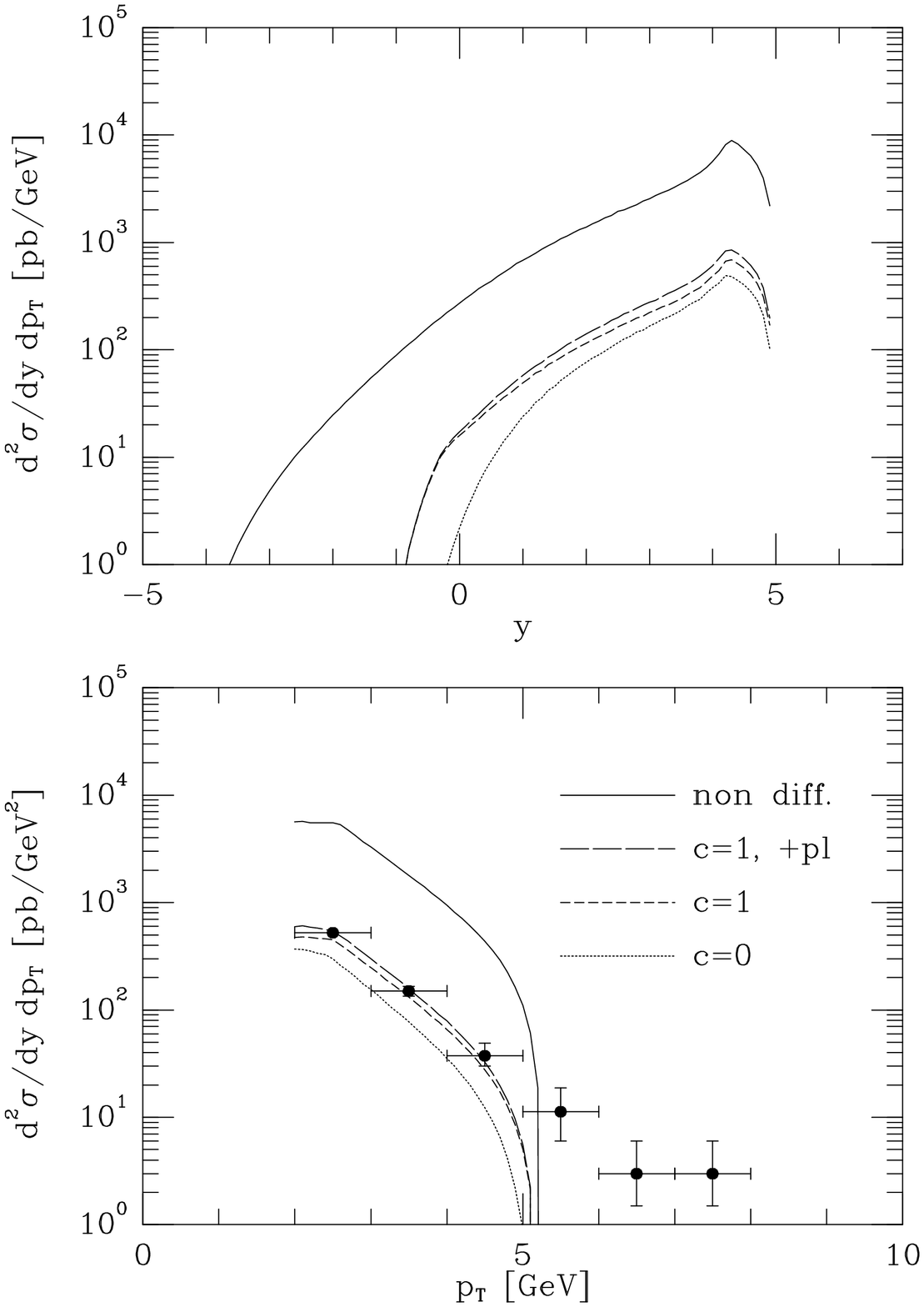}{120mm}{\em
The $y$ distribution and $p_T$ spectrum of the one jet cross
section in DIS, evaluated in the $\gamma^{*}p$ cms for fixed
$p_T=2\,\mbox{GeV}$ and $y=4$, respectibly. We impose
the experimental conditions $Q^2>10\,\mbox{GeV}^2$ and
$W^2>140^2\,\mbox{GeV}^2$ which correspond to the data points from
ZEUS \protect\cite{ZS063}.
}{fig8}

In DIS, the photon is always direct. The jet cross section
has been calculated for our model in the $\gamma^{*}p$ cms. Jets
with positive rapidity are travelling now in the photon direction.
With $p$, the four--momentum of the proton, $k$ the four--momentum of
the electron, and $q$, the momentum transfer to the photon,
we define the usual kinematic variables
$s=(p+k)^2$, $Q^2=-q^2$,
$W^2=(p+q)^2$, $x=Q^2/(2pq)$,
$y_e=pq/pk$, $z=p_T e^y/W$.
Finally, we set $\xi=x+p_T^2/(yz(1-z)s)$,
which is the fraction of energy
delivered from the proton to the subprocess (see \fref{fig1}).

The inclusive one--jet cross section is then given by
\begin{eqnarray}
\label{DIS}
  \frac{d^2\sigma}{dy\,dp_T}
  \hspace{-2mm}&=&\hspace{-2mm}
  \int_{a+b}^1 dy_e \int_{a/y_e}^{1-b/y_e}dx
  \frac{d^4\sigma}{dx\,dy_e\,dy\,dp_T}
  \quad,
\end{eqnarray}
where the kinematic bounds follow from the requirements
$Q^2 \ge Q^2_{min}$ and $W^2 \ge W^2_{min}$: $a=Q^2_{min}/s$,
$b=\max((2p_T\cosh y)^2,W^2_{min})/s$.

In the resolved pomeron case, we have
\begin{eqnarray}
   && \hspace{-10mm}
   \frac{d^4\sigma}{dx\,dy_e\,dy\,dp_T^2} \\
   \hspace{-4mm}&=&\hspace{-2mm}
   \sum_{bi} \int_\xi^{x_{I\hspace{-1mm}P}^{max}}
   \frac{dx_{I\hspace{-1mm}P}}{x_{I\hspace{-1mm}P}}
   G_{{I\hspace{-1mm}P}/p}(x_{I\hspace{-1mm}P})
   G_{b/{I\hspace{-1mm}P}}(\xi/x_{I\hspace{-1mm}P},Q^2)  \nonumber\\
   \hspace{-4mm}&&\hspace{-2mm}
   \frac{2\alpha_s \alpha^2 Q_i^2}{\xi x y_e^3 (1-z) s^2}
   \left\{ [1+(1-y_e)^2]\hat{h}_u+2(1-y_e)\hat{h}_l
   \right\} \hspace{2mm}.
   \nonumber
\end{eqnarray}

The contribution of the direct coupling is
\begin{eqnarray}
   && \hspace{-10mm}
   \frac{d^4\sigma}{dx\,dy_e\,dy\,dp_T^2} \\
   \hspace{-2mm}&=&\hspace{-2mm}
   \sum_{i}
   G_{{I\hspace{-1mm}P}/p}(x_{I\hspace{-1mm}P})
   \frac{2c^2 \alpha^2 Q_i^2}{4\pi\xi x y_e^3 (1-z) s^2} \\
   \hspace{-2mm}&&\hspace{-2mm}
   6 \left\{ [1+(1-y_e)^2]\hat{h}_u+2(1-y_e)\hat{h}_l
   \right\} \quad.
   \nonumber
\end{eqnarray}

The functions $\hat{h}_u\equiv\frac{1}{2}(\hat{h}_g+\hat{h}_l)$
and $\hat{h}_l$ depend on the Mandelstam variables
$\hat{s}+\hat{t}+\hat{u}=-Q^2$ of the subprocesses. For
$\gamma q \rightarrow q g$, we have
\begin{eqnarray}
   \hat{h}_g
   \hspace{-2mm}&=&\hspace{-2mm}
   \frac{4}{3}\left(-\frac{\hat{t}}{\hat{s}}-\frac{\hat{s}}{\hat{t}}
   +\frac{2Q^2\hat{u}}{\hat{s}\hat{t}}
   \right) \quad, \\
   \hat{h}_l
   \hspace{-2mm}&=&\hspace{-2mm}
   \frac{4}{3}\frac{-2Q^2\hat{t}}{(Q^2+\hat{s})^2} \quad,
\end{eqnarray}
while for $\gamma g\rightarrow q\overline{q}$, we find
\begin{eqnarray}
   \hat{h}_g
   \hspace{-2mm}&=&\hspace{-2mm}
   \frac{1}{2}\left(\frac{\hat{t}}{\hat{u}}+\frac{\hat{u}}{\hat{t}}
   -\frac{2Q^2\hat{s}}{\hat{t}\hat{u}}
   \right) \quad, \\
   \hat{h}_l
   \hspace{-2mm}&=&\hspace{-2mm}
   \frac{1}{2}\frac{4Q^2\hat{s}}{(Q^2+\hat{s})^2} \quad.
\end{eqnarray}

Like in the photoproduction case,
absolute ex\-peri\-men\-tal data for the
rapidity distribution or $p_T$ spectrum are not yet available to us.
In \fref{fig8}, we compare the shape of the $p_T$
spectrum with 1993 ZEUS data \cite{ZS063}. The experimental conditions
were $Q^2>10\,GeV^2$, $W>140\,GeV$ and $0.04<y_e<0.95$. No special case
of our model is favoured,since the slopes of direct and resolved
contribution are identical, but the shape can be approximately reproduced.

In our analysis, we did not consider particle production
or hadronization effects etc. The Monte--Carlo--programs
POMPYT by Bruni and Ingelman~\cite{BI} or RAPGAP by Jung~\cite{Ju}
have been developed for a wide class of pomeron models
and allow, together with other programs, the study of event
characteristics. They are widely used by the HERA collaborations
to interpret the large rapidity gap data.
\section{Conclusion}
In summary, we have studied the effect of a direct pomeron coupling
on diffractive jet production at HERA. The concept of pomeron
structure functions with DGLAP $Q^2$--evolution has to be enlarged
if the pomeron has a direct coupling to quarks. We have included
the additional point--like part in the analysis of diffractive
$F_2$ data and find consistency for the assumption of a
direct pomeron coupling to quarks. Some evidence for a direct
coupling has been found in the $p_T$ spectrum of photoproduction.
However, our analysis is model dependent and second,
except for the discussion of $\tilde{F}_2^D$ and $F_2^D$,
we have compared only the shapes and not the normalizations
of the photoproduction and DIS cross sections.
\begin{center}
{\large\bf Acknowledgements}
\end{center}
It is a pleasure to thank the conveners of the working group I$+$II
(Structures, Diffractive Interactions and Hadronic Final States)
and the organizing committee of DIS95. I am greatful to the
European Network {\it Physics at High Energy Colliders} for
invitation to the Second Meeting in Como.
%

\end{document}